\documentclass[conference]{IEEEtran}
\IEEEoverridecommandlockouts

\makeatletter                   
\def\mdseries@tt{m}             
\makeatother                    

\pdfoutput=1
\usepackage{microtype}
\usepackage{graphicx}
\usepackage{subcaption}
\usepackage{booktabs} 

\usepackage{hyperref}
\usepackage{amssymb} 
\usepackage[frozencache=true,cachedir=minted-cache]{minted}

\usepackage{amsmath}
\usepackage{mathtools}
\usepackage{amsthm}

\usepackage[capitalize,noabbrev]{cleveref}

\usepackage[textsize=tiny]{todonotes}

\usepackage[T1]{fontenc}
\usepackage{beramono}
\usepackage{tikz}
 \usetikzlibrary{positioning, arrows, automata, decorations.pathreplacing}
\usepackage{listings}
\lstdefinestyle{sharpc}{language=[Sharp]C}

\definecolor{dkgreen}{rgb}{0,0.6,0}
\definecolor{gray}{rgb}{0.5,0.5,0.5}
\definecolor{mauve}{rgb}{0.58,0,0.82}

\lstMakeShortInline{!}
\lstdefinestyle{sharpc}{language=[Sharp]C}
\usepackage{tikz}
\usetikzlibrary{positioning,calc}
\tikzset{onslide/.code args={<#1>#2}{%
  \only<#1>{\pgfkeysalso{#2}} 
}}

\makeatletter
\newenvironment{btHighlight}[1][]
{\begingroup\tikzset{bt@Highlight@par/.style={#1}}\begin{lrbox}{\@tempboxa}}
{\end{lrbox}\bt@HL@box[bt@Highlight@par]{\@tempboxa}\endgroup}

\newcommand\btHL[1][]{%
  \begin{btHighlight}[#1]\bgroup\aftergroup\bt@HL@endenv%
}
\def\bt@HL@endenv{%
  \end{btHighlight}%
  \egroup
}
\newcommand{\bt@HL@box}[2][]{%
  \tikz[#1]{%
    \pgfpathrectangle{\pgfpoint{1pt}{0pt}}{\pgfpoint{\wd #2}{\ht #2}}%
    \pgfusepath{use as bounding box}%
    \node[anchor=base west, fill=orange!30,outer sep=0pt,inner xsep=1pt, inner ysep=0pt, rounded corners=3pt, minimum height=\ht\strutbox+1pt,#1]{\raisebox{1pt}{\strut}\strut\usebox{#2}};
  }%
}
\makeatother

\usepackage{algorithm}
\usepackage{algorithmic}

\newcommand*\circled[1]{\tikz[baseline=(char.base)]{
            \node[shape=circle,draw,inner sep=0.25pt] (char) {#1};}}

\begin{document}
\title{RAPGen: An Approach for Fixing Code Inefficiencies in Zero-Shot}






\author{
\IEEEauthorblockN{Spandan Garg\textsuperscript{*}\thanks{\textsuperscript{*}Corresponding author}}
\IEEEauthorblockA{Microsoft Corporation\\
One Microsoft Way\\
Redmond, WA 98052, USA \\
spgarg@microsoft.com}
\and
\IEEEauthorblockN{Roshanak Zilouchian Moghaddam}
\IEEEauthorblockA{Microsoft Corporation\\
One Microsoft Way\\
Redmond, WA 98052, USA \\
rozilouc@microsoft.com}
\and
\IEEEauthorblockN{Neel Sundaresan}
\IEEEauthorblockA{Microsoft Corporation\\
One Microsoft Way\\
Redmond, WA 98052, USA \\
neels@microsoft.com}
}

\date{}
\maketitle
\begin{abstract}
Performance bugs are non-functional bugs that can even manifest in well-tested commercial products. Fixing these performance bugs is an important yet challenging problem. In this work, we address this challenge and present a new approach called Retrieval-Augmented Prompt Generation (RAPGen). Given a code snippet with a performance issue, RAPGen first retrieves a prompt instruction from a pre-constructed knowledge-base of previous performance bug fixes and then generates a prompt using the retrieved instruction. It then uses this prompt on a Large Language Model in zero-shot to generate a fix. We compare our approach with the various prompt variations and state of the art methods in the task of performance bug fixing. Our empirical evaluation shows that RAPGen can generate performance improvement suggestions equivalent or better than a developer in $\sim$60\% of the cases, getting $\sim$42\% of them verbatim, in an expert-verified dataset of past performance changes made by C\# developers. Furthermore, we conduct an in-the-wild evaluation to verify the model's effectiveness in practice by suggesting fixes to developers in a large software company. So far, we have shared performance fixes on 10 codebases that represent production services running in the cloud and 7 of the fixes have been accepted by the developers and integrated into the code. 

\end{abstract}


\thispagestyle{empty}
\pagestyle{plain}

\section{Introduction}
\label{submission}

Performance bugs are inefficient code snippets in software that can unnecessarily waste time and resources. 
Unlike functional bugs, performance bugs do not usually cause system failure. They tend to be harder to detect~\cite{attariyan2012x, dean2014perfscope, perfscope, catchmeifyoucan} and fix compared to functional bugs~\cite{caramelnistor, song2014oopsla} and are usually fixed by expert developers \cite{chen2019inferring} and often manifest through large inputs or specific execution configurations \cite{caramelnistor, olivo2015static}. 
Therefore, they can sometimes go undetected for a long periods of time~\cite{perfscope, catchmeifyoucan}. \
As a result, better tool support is needed to fix performance bugs, specially for novice developers. For such a tool to be applicable in practice, it should also support a wide-range of performance bugs. However, the existing fixing approaches usually target specific kinds of performance bugs, such as repeated computations \cite{memoization}, software misconfigurations \cite{misconfigurations},
loop inefficiencies \cite{caramelnistor}. The majority of these approaches are also rule-based analyzers, with high maintenance cost \cite{bielik2017learning}. Recently, to address this challenge, Garg et al. developed a deep-learning based performance fix tool called DeepDev-PERF~\cite{FSEPerf} that supports a variety of performance bugs and does not require maintaining explicit rules or an expert system. Our approach builds on that work by showing the cost of building and maintaining such a tool can be further improved through the use of prompt engineering as opposed to intensive fine-tuning.

\newcommand{\multilinecomment}[1]{}
\multilinecomment{
\begin{lstlisting}[linewidth=9cm, gobble=4, basicstyle=\tiny\ttfamily]
    public override void Undo(Params param) {
        foreach (Container obj in containers) {
            Beatmap copy = Beatmap.GenerateCopy(obj.objectData);
            `param.collections.Where(x => x.Type == copy.Type)`
                                   `.FirstOrDefault()?.SpawnObject(copy, out _);`
            if (obj is EventContainer e && e.eventData.IsRotationEvent)
                param.tracksManager.RefreshTracks();
        }
    }
    
\end{lstlisting}}

\begin{figure}[h]
\centering
\includegraphics[width=0.48\textwidth]{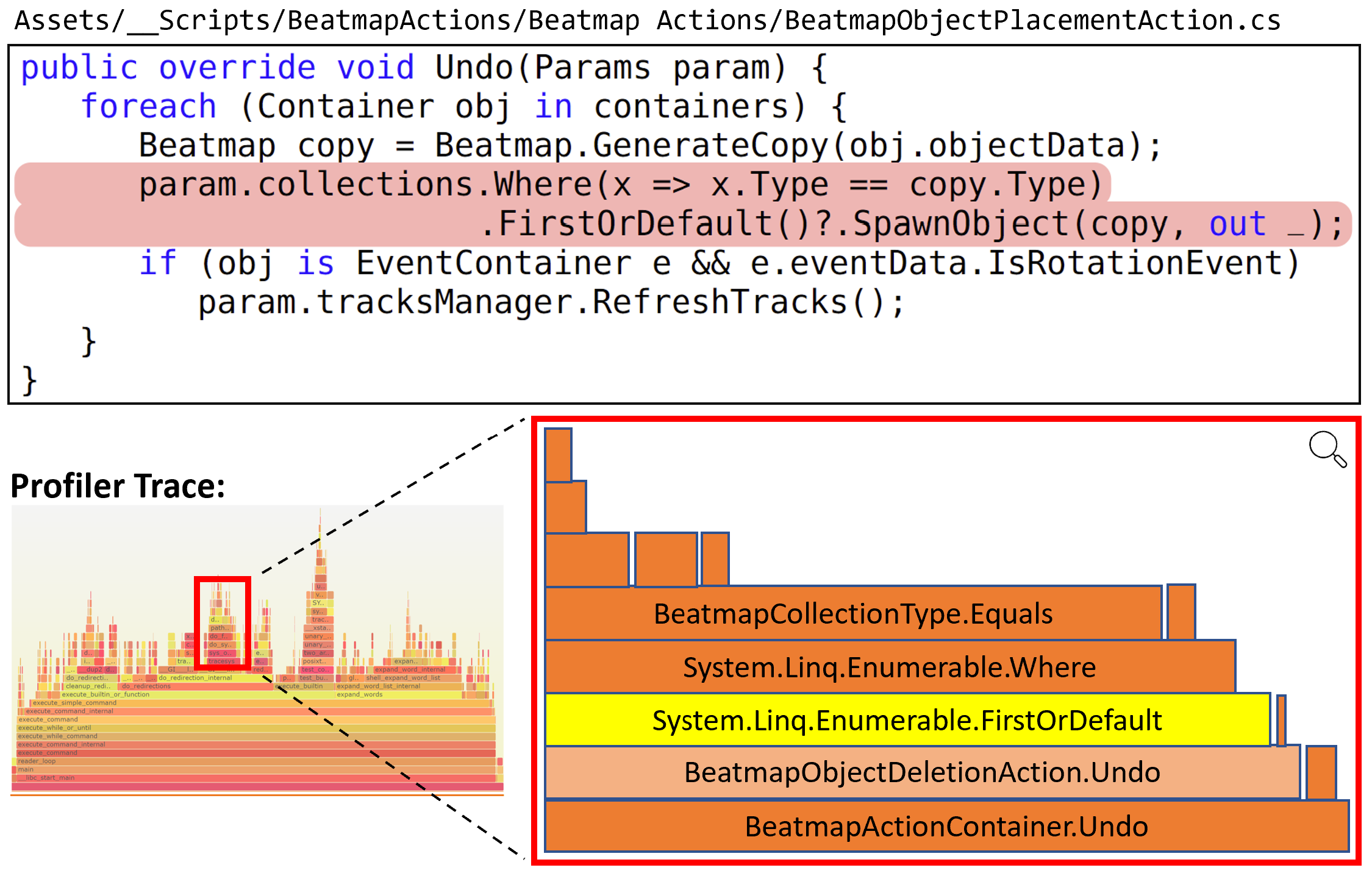}
\caption{A C\# code snippet with an expensive LINQ query (highlighted in red) from a performance bug fix commit on GitHub. LINQ tends to get misused by developers and can often lead to performance issues, such as the one above. 
In this case, the LINQ methods \texttt{\scriptsize{Where}} and \texttt{\scriptsize{FirstOrDefault}} are used to iterate over a collection to find all entries matching a predicate, when only the first match is needed and the search could potentially stop early. Depending on the size of the collection and how frequently this code gets invoked, this may become a performance hot-spot in an application. The screenshot below shows flamegraph corresponding to this application, with the relevant section highlighted. The call-stack shows the \texttt{\scriptsize{FirstOrDefault}} (its corresponding frame being highlighted in yellow) as being the most expensive line within the \texttt{\scriptsize{Undo}} method.}
\label{example_before}
\end{figure}
With the rising adoption of tools such as Github copilot~\cite{Codex} by developers, large language models are showing a promising success as productivity tools. Additionally, a promising paradigm for few-shot and zero-shot learning has emerged where a pre-trained large language model is adapted to various downstream tasks by conditioning it on natural language prompts. This paradigm has also been applied to a variety of tasks in Software Engineering, including the task of bug-fixing~\cite{SecurityCodex, FSEZeroShot}. 
Building upon this prior work, we present an approach called RAPGen that uses large language models (LLMs) in zero-shot, for the task of generating fixes for performance bugs using prompt engineering. 
Given a line of code that contains a performance bug (i.e. the buggy line), 
we compare this line with a pre-constructed knowledge-base to retrieve a prompt instruction that can be used to convey what change needs to be made to a LLM. The index contains patterns of buggy API usages extracted from a data-set of past performance fixes, along with an instruction explaining the fix transformation. Given a method with a performance issue and the corresponding buggy line within the method, we first retrieve an instruction from the knowledge-base and then construct an input prompt for the model, containing the buggy method, followed by the instructive comment and the method signature, coaxing the model to output the fixed version of the method. This type of zero-shot bug-fixing experience is desirable because (1) it avoids an expensive fine-tuning step, (2) alleviates the need to find a large high-quality labeled data-set of performance fixes for training such a model, and (3) fixes a wide-range of performance problems and could easily be extended to more bugs (by growing the number of commits used to build the knowledge-base) or other languages (by collecting relevant commits from other languages), without needing to train a separate model. 

\textbf{Evaluation and Effectiveness.} We evaluated our approach on the most recent performance bug fix data set used in the DeepDev-PERF study~\cite{FSEPerf}. Experimental results show that our approach is indeed able to generate correct suggestions for 60\% of the dataset, often within the very first suggestion, and out-performs DeepDev-PERF and other baselines for single method performance fixes. While designing an optimal prompt is still an open problem and remains elusive as ever, we also compare our approach with various prompt variations to demonstrate our design choices and our experiments with recent ideas such as reasoning-based prompting~\cite{chainofthought}. Furthermore, through our "In-the-wild" evaluation, we show real-world evidence that RAPGen generates suggestions that lead to actual performance improvements to real .NET service codebases within a large software company that are currently in production. These changes were validated and adopted by real developers.

\section{Motivating Example}
In this section, we provide an overview of RAPGen using a motivating example. Figure \ref{example_before} shows a C\# code method with a performance bug in the form of an expensive LINQ~\cite{pialorsi2007introducing} query (highlighted in red). The query iterates over a collection, finding all entries that match the predicate passed to the LINQ method \texttt{\small{Where}}, finally using \texttt{\small{FirstOrDefault}} to get the first such entry for further processing. While other languages have equivalent concepts, LINQ is specific to C\# and is known to inherent allocations associated with its use. As a result, LINQ usage on the application's hot-path can lead to excessive allocations, which can cause unwanted symptoms such as spikes in garbage collection (GC), depriving the application of CPU resources, and reducing its throughput. 

\begin{figure}[h]
\centering
\includegraphics[width=0.5\textwidth]{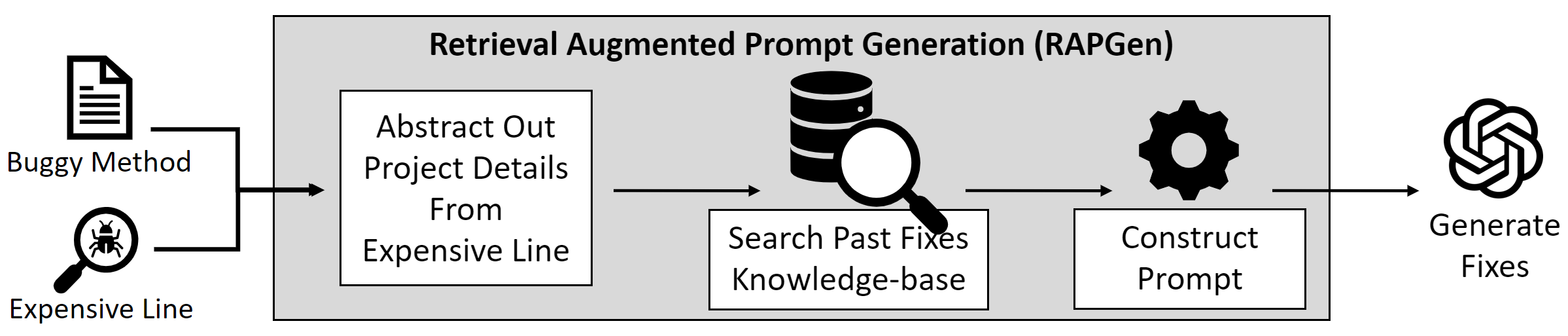}
\caption{The fix generation pipeline followed by RAPGen, showing how it is used at inference time to fix a performance issue given a buggy method and corresponding expensive line. }
\label{model_architecture}
\end{figure}

Figure \ref{model_architecture} shows the steps in our approach to generate fixes for a given buggy method and expensive line of code. To fix this code, we first compare the line of code containing the bug (that can be found easily using a profiler~\cite{app-insight-2019}), by abstracting out project-specific details and querying a pre-built knowledge-base containing past buggy API usages, which includes the buggy usages of \texttt{\small{FirstOrDefault}}. This gives us a prompt instruction in English language telling how this bug should be fixed, which we then use to construct the prompt as shown in Figure \ref{example_prompt}. The prompt includes the buggy method itself as a comment, followed by the prompt instruction, telling the model to re-write the preceding method without the \texttt{\small{FirstOrDefault}} call, implying that it should be replaced with cheaper alternatives, which in this case would be an inlined \texttt{\small{foreach}} loop. To further ensure that the the model completion contains the commented method, the prompt also includes the signature of the buggy method, followed by an open curly brace, at the end of prompt. This is to coax the model into outputting a completion of that same method. Figure \ref{example_after} shows the completion generated by the LLM to this prompt, which indeed replaces the LINQ query with an explicit \texttt{\small{foreach}} loop. 

\begin{figure}[h]
\centering
\begin{lstlisting}[linewidth=9cm, gobble=4, basicstyle=\footnotesize\ttfamily, commentstyle=\color{dkgreen},
]
    /*
    public override void Undo(Params param) {
        foreach (Container obj in containers) {
            Beatmap copy = Beatmap.GenerateCopy(obj.objectData);
            param.collections.Where(x => x.Type == copy.Type)
                                   .FirstOrDefault()?.SpawnObject(copy, out _);
            if (obj is EventContainer e && e.eventData.IsRotationEvent)
                param.tracksManager.RefreshTracks();
        }
    }
    */
    /* PERF: Rewrite the above method without FirstOrDefault. */
    public override void Undo(Params param) {

\end{lstlisting}
\caption{Model input prompt used to generate a fix for the method in Figure \ref{example_before}. The prompt consists of (i) the commented original buggy method, (ii) an instruction telling the model how to fix the issue, and (iii) the starting fragment of the buggy method.} 
\label{example_prompt}
\end{figure}

\begin{figure}[h]
\centering
\begin{lstlisting}[linewidth=9cm, gobble=4, basicstyle=\footnotesize\ttfamily, commentstyle=\color{dkgreen},
  ]
    public override void Undo(Params param) {
        foreach (Container obj in containers)
        {
            Beatmap copy = Beatmap.GenerateCopy(obj.objectData);
            @foreach (var collection in param.collections)@
            @{@
                @if (collection.Type == copy.Type)@
                @{@
                    @collection.SpawnObject(copy, out _);@
                    @if (obj is EventContainer e &&@ 
                         @e.eventData.IsRotationEvent)@ 
                        @param.tracksManager.RefreshTracks();@
                        
                    @break;@
                @}@
            @}@
        }
    }
    
\end{lstlisting}
\caption{Suggestion generated by the LLM when asked to complete the prompt in \ref{example_prompt}. The suggested fix is to unroll the LINQ query in favour of a \texttt{\scriptsize{foreach}} loop, which can stop early when a matching entry is found. This fix  closely matches the developer fix and saves other potential LINQ overheads such as the allocations and GC.}
\label{example_after}
\end{figure}

\section{Our Approach}
In this section, we explain the approach behind RAPGen and how we use it at inference time to fix performance bugs. The basis of our approach is building a code transformation knowledge-base with a code change specific retrieval algorithm, which we use to access the relevant prompt instruction for a given buggy code and construct the prompt.

\subsection{Code Transformation Knowledge-base}
\label{kb}
The first step of our proposed approach is to build a knowledge-base of performance code transformations and corresponding instructions, which can be used to coerce the language model into applying the same code transformation towards fixing a given bug. 
Below we describe how we collect the relevant data for building such a knowledge-base, as well as the processing steps needed to extract the necessary information to build this knowledge-base. 

\textbf{Data Collection}
\label{commits}
We first collect all repositories with $\geq$5 stars on GitHub, whose primary language is C\# and had a commit within the last 5 years. We crawl the commit history of the main branch in each of these repos. A commit typically contains a commit message and a changeset or diff representing the difference between current and previous version of files modified within that commit. From these repos, we collect the set of commits whose titles begin with "PERF:" or "[PERF]". This is a convention~\cite{commit} that's often used by developers to indicate that their commit contains primarily performance-related changes. This results in $\sim$1500 commits with clean code changes pertaining to performance.

\label{abstraction_algorithm}
\textbf{Extracting Performance Bug Patterns.} For each commit, we take the changeset, $D = [(f_{i}, f^\prime_{i})]_{i=1}^N$, where $f_{i}$ and $f^\prime_{i}$ are the before and after versions of the $i^{th}$ file modified in the commit. We parse the code files, $f_{i}$ and $f^\prime_{i}$, and extract all pairs of methods, $[(m_{i}, m^\prime_{i})]_{i=1}^M$, such that $m_{i}$ and $m^\prime_{i}$ are the before and after versions of a given method as modified by the commit, respectively.

\begin{algorithm}[tb]
   \caption{Abstract Impact Line}
   \label{abstraction_algorithm_code}
\begin{algorithmic}
\normalsize
   \STATE {\bfseries Input:} AST $node$, Array $ret$
   \IF{$IsLeaf(node)$}
       \IF{$IsKeyword(node)$ \OR $IsSyntax(node)$}
       \STATE $ret = Append(ret, node)$
       \ELSIF{$IsCommonIdentifier(node)$}
       \STATE $ret = Append(ret, node)$
       \ELSE
       \STATE $ret = Append(ret, Node_{\varnothing})$
       \ENDIF
   \ENDIF
   \FOR{$child \in node.Children$}
       \STATE $childArr = AbstractImpactLine(child)$
       \IF{$ContainsCommonIdentifier(childArr)$}
           \STATE $ret = Concatenate(ret, childArr)$
       \ELSE
           \STATE $ret = Append(ret, Node_{\varnothing})$
       \ENDIF
   \ENDFOR
\end{algorithmic}
\end{algorithm}

\multilinecomment{\begin{algorithm}[tb]
   \caption{Abstract Impact Line}
   \label{alg:example}
\begin{algorithmic}
   \STATE {\bfseries Input:} AST $node$, size $m$, Stack $s$
   \REPEAT
   \STATE Initialize $noChange = true$.
   \FOR{$i=1$ {\bfseries to} $m-1$}
   \IF{$x_i > x_{i+1}$}
   \STATE Swap $x_i$ and $x_{i+1}$
   \STATE $noChange = false$
   \ENDIF
   \ENDFOR
   \UNTIL{$noChange$ is $true$}
\end{algorithmic}
\end{algorithm}}

Given a method from the diff, $m_{i}$ with $K$ statements $[s_{j}]_{j=1}^K$, and its corresponding modified version, $m^\prime_{i}$ with $K^\prime$ statements $[s^\prime_{k}]_{k=1}^{K^\prime}$, we extract the statements $s_{j}$ that appear in $m_{i}$, but not in $m^\prime_{i}$. In other words, we find the set of statements, $S_d = \{s_{j} \mid 1 \leq j \leq K\text{ and }s_{j} \notin S_{m^\prime_{i}}\}$, where $S_{m^\prime_{i}}$ is the set of statements in ${m^\prime_{i}}$. We then use Algorithm \ref{abstraction_algorithm_code} to process all the statements in $S_d$. The goal of this algorithm is to remove any project-specific identifiers, such as variable names or user-defined function names, replacing them with placeholders. This yields a set of $L$ abstracted performance bug patterns $[p_{i}]_{i=1}^L$, where $p_{i} = f(s_{j})$ and $s_{j} \in S_d$.  

This leaves us templates of modified usages of common C\# API, containing only tokens corresponding to language syntax/keywords and common literals/identifiers in C\#.  
This type of statement abstraction allows us to pick up patterns in changes that get repeated across projects and may correspond to common performance issues. To further avoid any project-specific change patterns, we only keep patterns occur in at least 2 projects. 

Below is an example of one such transformation made by the abstraction algorithm:
\begin{minted}
[fontsize=\footnotesize,escapeinside=||, mathescape=True]{C}
    Foo().|\bfseries{Where}|(x => x.Bar()).|\bfseries{FirstOrDefault}|(); 
                        |$\downarrow$|
        |<$Node_{\varnothing}$>|.|\bfseries{Where}|(|<$Node_{\varnothing}$>|).|\bfseries{FirstOrDefault}|();
\end{minted}

The first line contains a possible statement modified as part of a commit. The abstraction algorithm traverses the AST corresponding to this statement and replaces all project-specific identifiers/literals (e.g. \texttt{\small{Foo}} and \texttt{\small{Bar}}) or any subtrees, none of whose descendants are known common identifiers in C\#, with a placeholder node (<$Node_{\varnothing}$>). We then traverse the code corresponding to this abstracted tree to get the code with project-specific details abstracted out. In the above example, the identifiers \texttt{\small{Where}} and \texttt{\small{FirstOrDefault}} are retained because these are frequently used .NET library functions in C\# projects. The above pattern specifically, is an example (Figure \ref{example_before}) of a known inefficient usage pattern where functions, \texttt{\small{Where}} and \texttt{\small{FirstOrDefault}}, are chained together inefficiently. This could be re-written to be more optimal by condensing the query to remove the \texttt{\small{Where}} call or simply unrolling the query into an explicit loop.

\textbf{Creating a Code Transformation Instruction} Now for each extracted performance bug pattern $p_{i}$, we find all the $K$ before-after pairs of methods, $[(m_k, m^\prime_{k})]$ from performance improvement commits, $C_{PERF}$, such that the buggy statement, $s$, which is the concrete version of $p_{i}$, is contained within $m_k$ and not in $m^\prime_{k}$.

For each such before-after pair, we would like to create an instruction $t_{k}$ that explains the code transformation taking place as part of the change. To do this, we find $(I_{m_k}, I_{m^\prime_{k}})$, which are the sets of common C\# identifiers added and/or removed as part of this code transformation, respectively. 

To extract the identifiers, we compare the sets of statements in $m$ and $m^\prime$, i.e. $S_{m}$ and $S_{m^\prime}$, respectively. For every statement in the diff i.e. statements that appear in $m$ and not in $m^\prime$, or ones that appear in $m^\prime$ and not in $m$, in other words, ${(S_m \setminus S_{m^\prime})} \cup {(S_{m^\prime} \setminus S_m)}$, we extract any common C\# identifiers used within these lines. These common literals/identifiers that correspond to popular library functions in C\# and can be found by looking for literals and identifiers that repeat across multiple projects. We consider the set of common C\# identifiers found in ${(S_m \setminus S_{m^\prime})}$ and ${(S_{m^\prime} \setminus S_m)}$. We then remove any identifiers that are found to be common among the two sets. We call the resulting disjoint sets of identifiers $I_{m}$ and $I_{m^\prime}$, respectively. %


To create $t_{k}$, we focus on the following code fix transformations that involve adding or removing identifiers (or APIs):  
\begin{itemize}
\item \textbf{Identifier Addition}: In this case, $I_{m_k} = \emptyset$ and $I_{m^\prime_{k}} \neq \emptyset$ and $t_{k} = $ \textit{"PERF: Rewrite the above method with <X>."}. This represents an instruction to re-write the code with some additional identifiers (or APIs) needed to resolve the performance issue, which, in terms of code-changes, may be a new function call or an additional parameter added to an existing call, etc.
 \item  \textbf{Identifier Removal}: In this case, $I_{m_k} \neq \emptyset$ and $I_{m^\prime_{k}} = \emptyset$ and $t_{k} = $ \textit{"PERF: Rewrite the above method without <X>."}. This represents an instruction to remove the usage of some identifiers (or APIs), which may correspond to removal of the use of some inefficient APIs, within the code. The example in Figure \ref{example_prompt} falls within this category.
\item \textbf{Identifier Addition and Removal}: In this case, $I_{m_k} \neq \emptyset$ and $I_{m^\prime_{k}} \neq \emptyset$ and $t_{k} = $ \textit{"PERF: Use <X> instead of <Y> in the above method."}. This instruction is meant to represent the case where the model should replace one set of identifiers (or APIs) with another, i.e., situations where one set of API must be swapped out for another, which are more appropriate for the situation. 
\end{itemize}

In each case, we replace the placeholders in the template 
$t_k$, i.e. \texttt{\scriptsize{<X>}} and \texttt{\scriptsize{<Y>}}, with a string containing the comma-delimited list of identifiers in $I_{m}\text{ and }I_{m^\prime}$, respectively.

\label{index}
\textbf{Knowledge-base Creation} 
Finally, taking all the buggy usage patterns and their corresponding before-after pairs and instructions, we create a knowledge-base to store all the $(p, (m, m^\prime, t))$, where $p$ serves as the lookup key to find the set of triples, $(m, m^\prime, t)$, where $m$ and $m^\prime$ are the before-after methods, respectively, and $t$ is the derived instruction from this particular before-after pair. This knowledge-base is essentially a mapping between abstracted performance bug patterns and instructions on appropriate code transformations to fix those bugs. In the following section, we describe our retrieval process, to fetch the right instruction from the knowledge-base.

\subsection{Generating the Prompt}
\label{generate_prompt}

\multilinecomment{\begin{algorithm}[tb]
   \caption{Extract Method Of Interest}
   \label{method_extraction_algorithm_code}
\begin{algorithmic}
   \STATE {\bfseries Input:} AST $node$, Array $ret$
   \IF{$node.IsLeaf()$}
       \STATE $noChange = false$
       \IF{$node.IsKeyword()$ \OR $node.IsSyntax()$}
       \STATE $ret.Append(node)$
       \ELSIF{$node.IsCommonIdentifier()$}
       \STATE $ret.Append(node)$
       \ELSE
       \STATE $ret.Append(Node_{\varnothing})$
       \ENDIF
   \ENDIF
   \FOR{$child \in node.Children$}
       \STATE $childArr = AbstractImpactLine(child)$
       \IF{$ContainsCommonIdentifier(childArr)$}
           \STATE $ret.Concatenate(childArr)$
       \ELSE
           \STATE $ret.Concatenate(Node_{\varnothing})$
       \ENDIF
   \ENDFOR
\end{algorithmic}
\end{algorithm}}
We are given buggy method, $m_b$, containing an expensive line of code, $l$, which causes the performance bug. 
We first abstract the expensive line using Algorithm \ref{abstraction_algorithm_code} and get the abstracted pattern $p_b$. We then perform a lookup in our knowledge-base 
using the $p_b$, as the query. This results in a list of $K_p$ code change and instruction triples, i.e. $[(m_k, m^\prime_k, t_k)]_{k=1}^{K_p}$, where $m_k$ and $m^\prime_k$ are the before and after versions of a method, where a similar buggy line was fixed and $t_k$ is the code transformation instruction constructed based on changes made to the identifiers in the before-after pair. Since we want to find the entry that's closest to $m_b$, we use a code-search technique ~\cite{luan2019aroma} to rank each before method, $m_i$, based on similarity to $m_b$. We take the entry corresponding to the most similar before method, $(m_j, m^\prime_j, t_j)$ and construct the prompt using the retrieved code transformation instruction $t_{j}$. 


Now we fill the template in Figure \ref{instruction_prompt_template} to build the final prompt, which will be given to the model. We include the buggy method, $m_b$, itself surrounded by C-style comments, followed by $t_j$ and the signature of $m_b$ and an open curly-brace, which in C\# indicates the start of the method body.

\begin{figure}[h]
\centering
\begin{lstlisting}[linewidth=10cm, basicstyle=\footnotesize\ttfamily, numbers=none, %xleftmargin=.28\textwidth, commentstyle=\color{dkgreen},frame=single
]
    /* 
    <Commented Buggy Method> 
    */
    /* PERF: <Prompt Message> */
    <Method Signature> {
\end{lstlisting}
\caption{High level prompt template followed by the prompts generated by RAPGen. They consist of the buggy method itself, followed by the prompt instruction retrieved from the KB using the buggy line within the buggy method and finally the signature of the method itself, proceeded by an open curly brace.}
\label{instruction_prompt_template}
\end{figure}

Figure \ref{example_prompt} shows a concrete example of one such prompt following the above template.

\multilinecomment{\definecolor{forestgreen}{rgb}{0.13, 0.54, 0.13}
\begin{table}[htbp]
    \centering
    \tiny
    \caption{\normalfont Prompt Message Templates.}
    \label{prompt_table}
    \begin{tabular}{l l l l l}
        ID & \textbf{Prompt Message Template Body} & \textbf{Category Fixed}
        \\\hline\hline
        1 & \textcolor{forestgreen}{\texttt{\tiny{PERF: Rewrite the above method with <X>.}}} &  \circled{1}, \circled{3}\\\hline
        2 & \textcolor{forestgreen}{\texttt{\tiny{PERF: Rewrite the above method without <X>.}}} &  \circled{1}, \circled{2}\\\hline
        3 & \textcolor{forestgreen}{\texttt{\tiny{PERF: Replace <X> with <Y> in the above code.}}} &  \circled{2}\\\hline
        4 & \textcolor{forestgreen}{\texttt{\tiny{PERF: <X> is on the hot-path. We can do better by ... }}} &  \circled{2}\\
    \end{tabular}
\end{table}}

\subsection{Generating Fixes}
Generating the fix itself is relatively simple. We use the constructed prompts to generate completions using the \textit{gpt-3.5-turbo}~\cite{openai_models} model through OpenAI's REST API. Since it is difficult to tell the model when to stop, it will likely go on outputting code even after it has generated the fixed method. We parse out the completion corresponding to the method by extracting the code until curly braces are balanced and discard the rest of the output.

\section{Empirical Evaluation}
In this section, we present our empirical evaluation we conducted on understanding our prompt design choices and assessing the effectiveness of our approach. We first describe our experimental setup, i.e. our dataset, metrics and baselines for our empirical evaluation.

\subsection{Experimental Setup}
\textbf{Dataset} We leverage the dataset containing past performance fixes from the DeepDev-PERF study~\cite{FSEPerf}. This dataset consists of 132 instances of performance fixes, which were manually reviewed and confirmed to be performance-related changes by performance experts after examining a broad set of C\# commits on GitHub. Furthermore, it was shown that this dataset covers a wide-range of performance bugs. Therefore, a good accuracy over this dataset would demonstrate the effectiveness of a model for this task.
Each example in this dataset is of the form $(m, m^\prime)$, where $m$ and $m^\prime$ are the before and after versions, respectively. These are also mostly single method performance improvement fixes, making them ideal for our scenario.  

\textbf{Adding Bug Localization Information} Our approach assumes that bug localization information, i.e. the expensive line, is included as part of the input. In practice, one can leverage performance bug localization using profiling data to find the expensive line \cite{garg2021perflens}. However, for the purpose of evaluating our approach, we used a heuristic to localize bugs in the above dataset. We examine the diff between the before and after methods, $m$ and $m^\prime$ and take the first line $l_b \in m$ that does not match the corresponding line in $m^\prime$, when compared in order, as the buggy line. 

\textbf{Inference Hyperparameters}
We generate 100 suggestions/fixes for each example in the dataset, using a relatively large temperature of 0.7 to achieve sufficient variety within the fixes generated. We also use a max token limit of 1024 tokens, so that the model has enough token bandwidth to generate the entire method for longer suggestions. Finally, we take the model output and parse out the fix as described in the previous section.



\textbf{Automated Metrics} We use the following three automated metrics for measuring the effectiveness of our approach: 
\begin{itemize}
    \item \textbf{Verbatim Match \%}: This metric considers a prediction to be correct only if the generated fix exactly matches the human fix in the test set. This metric usually presents a lower bound as two pieces of code can be equivalent, despite not matching verbatim. This metric is also sensitive to identifier names and order of code statements. 
    \item \textbf{Abstracted Match \%}: In this metric, we abstract out variable names using a placeholder name (such as "VAR\_\{i\}", e.g. "VAR\_0", "VAR\_{1}", "VAR\_{2}" ...), where $i$ is determined based on the order, in which we encounter the variables during a traversal of the AST of the method. Using the traversal order, ensures that two code snippets with the same control flow and no major changes except variable names, will have the same abstracted variable names. But, even this presents us with a lower bound on the approach's true capabilities in fixing bugs as a fix may have different control-flow from ground truth, but still be correct.
    \item \textbf{CodeBLEU}: Our final metric is CodeBLEU~\cite{Ren2020CodeBLEUAM}, which is a variation of the popular NLP metric, BLEU~\cite{Papineni2002BleuAM}. In addition to n-gram matching, it also takes code characteristics like abstract syntax tree (AST) and data-flow similarity into account, when comparing programs. Similar to the DeepDev-PERF study, we use the hyper-parameters that had the highest correlation to human scores in the CodeBLEU study, i.e. $\alpha, \beta, \gamma, \delta = 0.1, 0.1, 0.4, 0.4$.
\end{itemize}

\textbf{Human Evaluation Metrics} In addition to the above automated metrics, we also conduct a human evaluation of the suggestions with two performance experts (not on the author list). We use the following metric:
\begin{itemize}
    \item \textbf{Closest Match Top-K Accuracy \%} Due to there being too many suggestions to go through one by one, we use the state of the art code search technique, Aroma~\cite{luan2019aroma}, to find the suggestion closest to the ground truth. The most similar suggestion is shown to two performance experts, who are then asked whether they consider the model suggestion to be equivalent or better that the developer fix in terms of performance. We consider a suggestion to be correct, if it is found to be correct by both experts. This allows us to get a lower-bound on the model's actual Top-K accuracy, if one were to exhaustively go through every suggestion. 
\end{itemize}

\makeatletter
\newcommand{\srcsize}{\@setfontsize{\srcsize}{5pt}{5pt}}
\makeatother
\begin{figure*}
  \begin{subfigure}{0.5\textwidth}
    \begin{lstlisting}[linewidth=6cm, basicstyle=\footnotesize\ttfamily, numbers=none, commentstyle=\color{dkgreen},]
    /* 
    <Commented Buggy Method> 
    */
    /* PERF: Improve performance of the above method. */
    <Buggy Method Signature> {
    \end{lstlisting}
    \caption{Static Prompt Template} \label{fig:static}
  \end{subfigure}%
  \hspace*{\fill}   
  \begin{subfigure}{0.5\textwidth}
    \begin{lstlisting}[linewidth=6cm, basicstyle=\footnotesize\ttfamily, numbers=none, commentstyle=\color{dkgreen},]
/* 
<Commented Retrieved Buggy Method> 
*/
<Retrieved Fixed Method> 
/* 
<Commented Buggy Method> 
*/
<Buggy Method Signature> {
    \end{lstlisting}
    \caption{Retrieval-based One-Shot Prompt Template} \label{fig:few_shot}
  \end{subfigure}%
  \hspace*{\fill}   

  \begin{subfigure}{0.5\textwidth}
    \begin{lstlisting}[linewidth=5cm, basicstyle=\footnotesize\ttfamily, numbers=none, commentstyle=\color{dkgreen},]
/* 
<Commented Buggy Method> 
*/
/* PERF: <X> is on the hot-path in the above method. We can do better by ...
    \end{lstlisting}
    \caption{Reasoning-based Prompt Template} \label{fig:reasoning}
  \end{subfigure}

\caption{Prompt templates for various prompt variations we tried in our evaluation.} \label{prompt_templates}
\end{figure*}


\textbf{Prompt Variants}
To better understand our prompt design choices, we investigate the following prompt variations in addition to RAPGen:
\begin{itemize}
    \item \textit{Static Prompt Message.} This prompt follows the template in Figure \ref{fig:static}. We first add the buggy method, $m_b$ itself as a C-style comment, followed by a static comment expressing an intent to improve performance and, finally, the signature of the buggy method with an open curly brace. The purpose behind including this prompt is to contrast with our approach where we provide a prompt message containing information regarding where the bug is and hints on how to fix it, to show whether providing the model with additional information regarding the bug helps it come up with more fixes.

    \item \textit{Retrieval-based One-shot Prompt.} This prompt follows the structure in Figure \ref{fig:few_shot}. We first provide the model with the retrieved before-after pair, $(m_j, m_j^\prime)$ (from Section \ref{generate_prompt}) where $m_j$ contains a line, $l_{m_j}$, whose abstracted version matches the abstracted buggy line, $l_b$ in $m_b$. Similar to RAPGen prompts, we then provide the model with the buggy method as a comment, followed by the signature of the buggy method, coaxing it into outputting a fixed version of that method. Our reason for including this prompt is to show the impact of providing a raw fix example, as opposed to giving the model the instruction derived from the same code-change. 
    
    \item \textit{Reasoning-based Prompt Message.} Recent work on prompt engineering suggests letting LLMs generate an intermediate reasoning about a problem can greatly improve performance on a given task~\cite{chainofthought}. 
    As our final variant, we designed a reasoning-based prompt that follows the structure in Figure \ref{fig:reasoning}. This prompt also includes the buggy method, $m_b$, followed by an incomplete comment telling the model what call is being expensive, and coaxing the model to reason about the problem and then generate a possible fix. The main idea is that we first get the model to reason about the problem and come up with a description of the possible fix by completing the prompt message itself, before it generates the fixed code. We do this by giving the model the buggy method, followed by the prompt message. We let the model complete the rest of the comment (i.e. continuing in place of the ellipsis) and then have it generate a fix as well. The prompt follows structure in Figure \ref{fig:reasoning}.
\end{itemize}

\subsection{Comparing with Other Prompt Variants}
Table \ref{automated_eval_table} shows the results of our automated and human evaluation metrics, achieved by RAPGen and other prompt variants. Our approach is able to achieve a $\sim$42\% verbatim match and a $\sim$41\% abstracted verbatim match score over the dataset, which is higher than any of the other prompt variants. We also see that in the human evaluation, our approach is able to suggest valid improvements for up to $\sim$63\% suggestions within just 100 sampling attempts, higher than all our other prompt variant. We discuss these results and possible reasons behind them in more detail below.

\begin{table*}[t]
    \centering
    \scriptsize
    \caption{\normalfont Summary of the results of our approach against other prompt variants.}
    \label{automated_eval_table}
    \begin{tabular}{l| c| c| c | c c c}
         Model & Verbatim Match \% & Abstracted Match \% & CodeBLEU & \multicolumn{3}{c}{Closest Match Top-K Accuracy \%}\\
         & & & & 1 & 10 & 100\\\hline
        Static Prompt & 26.5 & 27.3 & 67.6 & 4.5  & 13.6 & 33.3\\\hline
        Retrieval-based One-shot Prompt & 28.8 & 29.5 & 69.7 & 6.1  & 14.4 & 37.9\\\hline
        Reasoning-based Prompt & 21.2 & 31.1 & 53.6 & 1.5  & 9.1 & 43.2\\\hline
        RAPGen & \textbf{42.4} & \textbf{45.5} & \textbf{72.0} & \textbf{17.4}  & \textbf{28.8} & \textbf{60.6}\\\hline
    \end{tabular}
\end{table*}

\begin{figure}[h]
\centering
\includegraphics[width=0.45\textwidth]{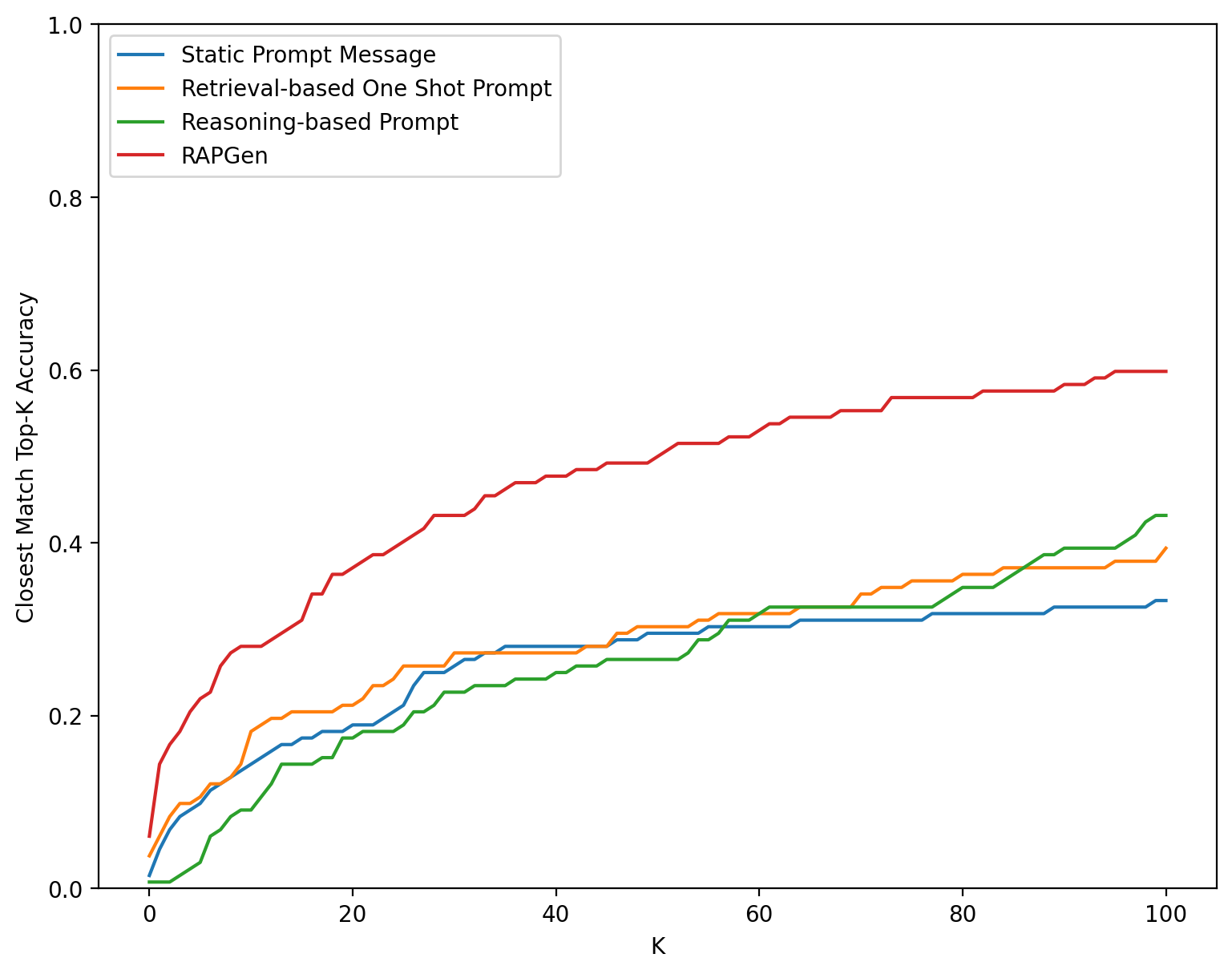}
\caption{\normalfont Closest Match Top-K Accuracy plot of our prompts on the test set. We can see that RAPGen achieves the best score among the 4 prompt variants.}
\label{top_k_plot}
\end{figure}

\begin{table*}[t]
    \centering
    \footnotesize
    \caption{\normalfont Summary of the results of our approach against other models.}
    \label{automated_eval_table_2}
    \begin{tabular}{l| c| c| c | c c c c}
         Model & Verbatim Match \% & Abstracted Match \% & CodeBLEU  & \multicolumn{4}{c}{Closest Match Top-K Accuracy \%}\\
         &&&&  1 & 10 & 100 & 500  \\\hline
        DeepDev-PERF & 34.3 & 37.3 & 70.7 & 8.3 & 18.2 & 34.1 & 53.0\\\hline
        RAPGen & \textbf{42.4} & \textbf{45.5} & \textbf{72.0} & \textbf{17.4}  & \textbf{28.8} & \textbf{60.6} & - \\\hline
    \end{tabular}
\end{table*}

\begin{figure}[!ht]
    \centering
        \centering
\begin{lstlisting}[linewidth=8cm, gobble=4, basicstyle=\footnotesize\ttfamily, commentstyle=\color{dkgreen},frame=single
  ]
    /*
    public IServiceProvider CreateContainer(ShellSettings settings)
    {
        // ...
        foreach (var dependency in blueprint.Dependencies)
        {
            foreach (var interfaceType in dependency.GetInterfaces())
            {
                _logger.LogDebug("Type: {0}, Interface Type: {1}", dependency.Type, interfaceType);
        // ...
    }
    */
    /* PERF: LogDebug is on the hot-path in the above method. We can do better by ...
\end{lstlisting}
        \centering
\begin{lstlisting}[linewidth=8cm, gobble=4,escapeinside=||, mathescape=True, basicstyle=\footnotesize\ttfamily, commentstyle=\color{dkgreen},frame=single]
    |\textcolor{dkgreen}{using LogDebug only when logging is enabled. */}|
    public IServiceProvider CreateContainer(ShellSettings settings)
    {
        foreach (var dependency in blueprint.Dependencies)
        {
            foreach (var interfaceType in dependency.GetInterfaces())
            {
                @if(_logger.IsEnabled(LogLevel.Debug))@
                @{@
                   _logger.LogDebug("Type: {0}, Interface Type: {1}", dependency.Type, interfaceType);
                @}@
                // ...
    }
\end{lstlisting}
\caption{\label{reasoning_prompt}A relatively high-level suggestion generated by the Reasoning-based prompt. The code on top (simplified) contains the commented out the buggy method, followed by an incomplete comment letting the model know about where the problem is, which, in this case, happens to be excessive calls to \texttt{\scriptsize{LogDebug}}. The suggestion at the bottom finishes the incomplete the comment with a message, showing that the model did indeed reason about the problem presented and came up with a possible fix. The suggestion that follows wraps the \texttt{\scriptsize{LogDebug}} calls with a \texttt{\scriptsize{Logger.IsEnabled}} checks. This was also the fix made by the developer.}
\end{figure}

\begin{figure}[!ht]
\centering
\includegraphics[width=0.47\textwidth]{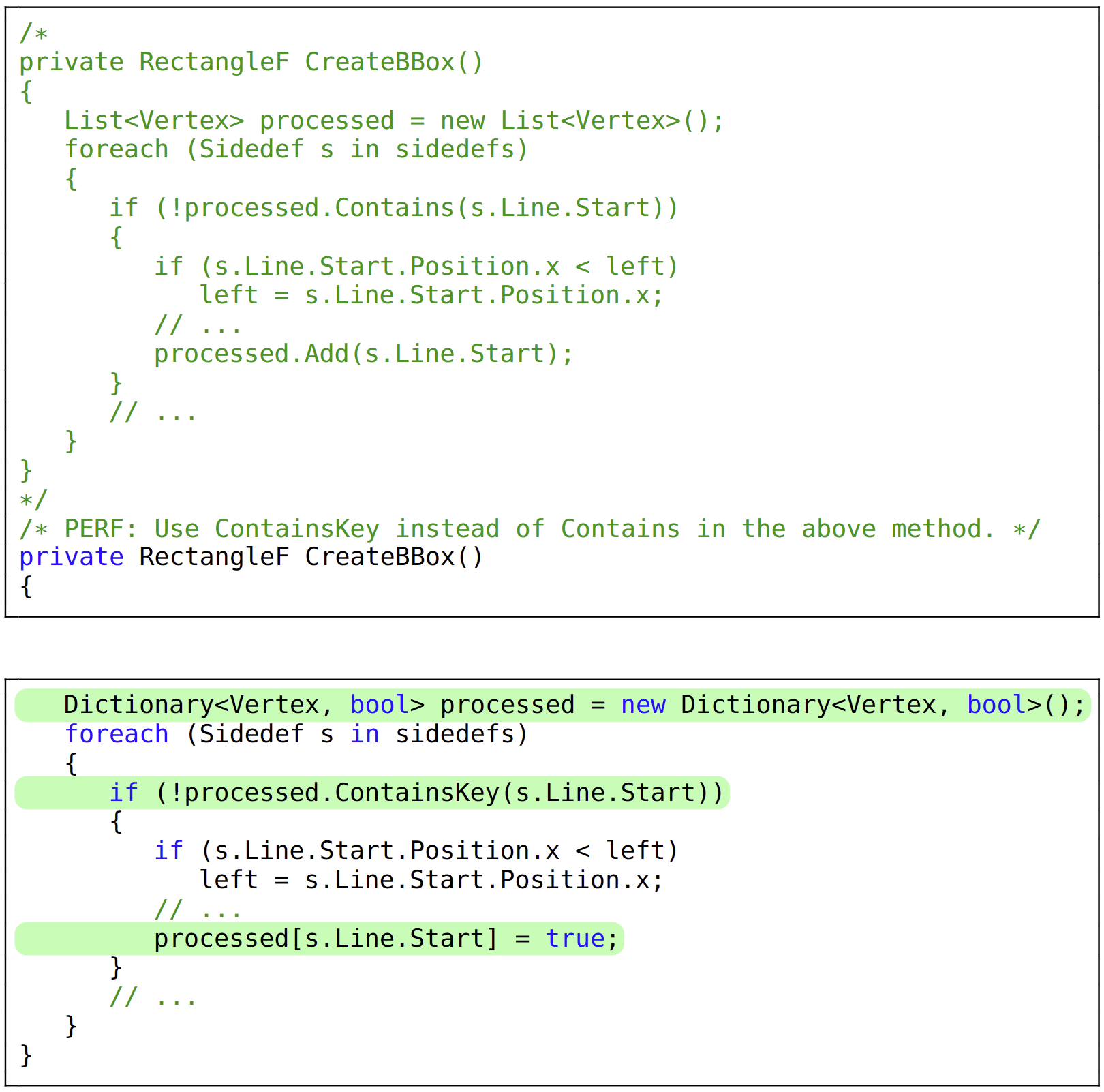}
\caption{\label{instructive_prompt_change}An API change suggested by RAPGen for one of the examples in the testset. The code on the top (shortened for brevity) shows the commented out buggy method with performance issues caused by the use of \texttt{\scriptsize{List.Contains()}} (a O(N) search) within a for-loop. The instruction retrieved from the knowledge-base is to replace \texttt{\scriptsize{List.Contains()}} with a \texttt{\scriptsize{Dictionary.ContainsKey()}} (an O(1) lookup). RAPGen prompt also includes the signature of the method, followed by an open curly brace to coax the model into outputting the same method. 
The second code snippet shows the completion generated by the model for this prompt, which is indeed to replace the \texttt{\scriptsize{List}} instansiation with that of a \texttt{\scriptsize{Dictionary}} and updating the calls within the loop to use \texttt{\scriptsize{ContainsKey}}. This change was also semantically equivalent to the fix made by the developer.}
\end{figure}

\textbf{Effect of adding instruction conveying the bug location and possible fix.} We first compare the results of the RAPGen's prompts with the Static prompts (Figure \ref{fig:static}). These two prompts are almost identical except for the prompt message following the commented buggy method. RAPGen's prompt includes a fix instruction based on past fixes to bugs of similar nature, which tells the model what function call within the method body is expensive, as well as a possible fix. On the other hand, the Static Prompt uses the same generic fix instruction for each bug. We see that RAPGen performs significantly better in terms of both the number of exact matches it finds,  $>$10\% more, as well as, achieves a higher Top-K score (by $>$25\%) compared to static prompts. Figure \ref{instructive_prompt_change} shows an example of a case where RAPGen was able to find a fix, whereas the Static prompt wasn't. Since this was a relatively long method, which required a non-trivial fix consisting of a data-structure change, Static prompt fails to come up with a fix. Whereas, RAPGen prompt is guided by the prompt instruction based on past fixes, which is better able to guide the model to the right fix. Intuitively, this makes sense because RAPGen's prompt message includes information about the location of the bug as well. 

\begin{figure}
\centering
\includegraphics[width=0.48\textwidth]{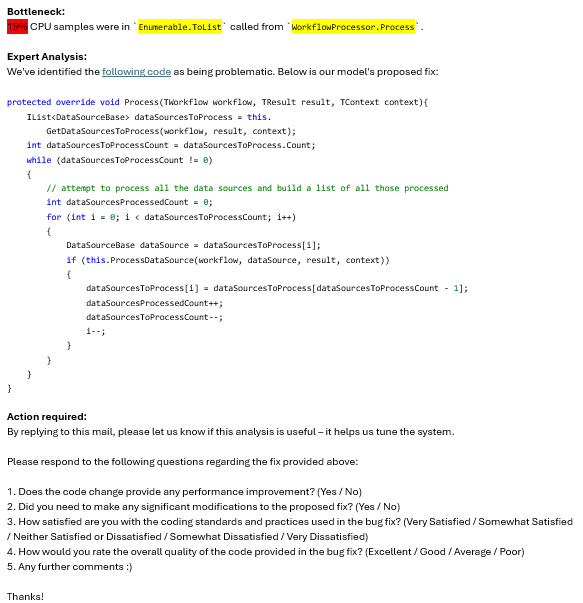}
\caption{An email sent as part of our "In-the-wild" Evaluation. The email contains a fix for a performance bottleneck found using a commercial performance bug detection service in an application. Along with the fix, we also include information regarding the bottleneck, i.e. the bottleneck method and the calling parent method (both highlighted in yellow) and current CPU usage of the bottleneck. At the end of the email, we ask the users to answer a few questions regarding the fix (Section \ref{survey_questions}).}
\label{email}
\end{figure}

\textbf{Effect of extracting instruction from retrieved example.} Prior work has shown that providing an example in a few shot setting usually results in better performance in a downstream task than providing an instruction\cite{brown2020language}. It likely has less to do with the example but more with specificity of the instructions
The Retrieval-based One-Shot prompt gives the model the same retrieved before-after pair that the instruction is derived from. However, comparing the results of RAPGen with the One-Shot prompt, we can see that going the extra step of extracting an instruction from the retrieved before-after pair improves the results significantly. We believe that this may be because, given the raw before-after method pair, the model has to perform the diff implicitly and infer the instruction by itself, before applying the change to the unseen method. It is possible for it misunderstand the change and make mistakes in understanding the nature of the fix. We see that our prompts find more exact matches (by $>$10\%) and achieves a higher Top-K accuracy score (by $>$20\%), compared to retrieval prompts. For the example shown in Figure \ref{instructive_prompt_change}, both the One Shot and RAPGen prompts were able to generate the right fix in our evaluation. However, the One Shot prompt took almost twice as many attempts before arriving at the right fix, compared to RAPGen. The likely reason RAPGen is able to generate a correct suggestion more quickly may be due to a possible fix being readily available as part of the instruction, without the model having to perform in-context learning from a raw code change example and infer the underlying transform.

\textbf{Effect of letting the model reason about the problem on its own.} Both Reasoning-based and RAPGen's prompts include information about where the buggy line is within the method. The main difference between them is that we do not provide the model with any hints on how to fix the bug in Reasoning-based prompts and instead allow it to reason about the problem and come up with a possible fix itself. During the human evaluation, we noticed that with such a reasoning-based prompt, the model is more likely to come up with high-level changes (introducing caching, memoization, adding a fast-path, etc.), modifying the control flow or the algorithm itself rather than an API-level fix. Figure \ref{reasoning_prompt} shows an example of one such change suggested for a reasoning based prompt, during our evaluation. Since high-level changes can be more complex to make, the model is more likely to create an incorrect or over-complicated fix. Therefore, it cannot arrive at the correct fix right away resulting in a flatter plot in the Figure \ref{top_k_plot}, despite given the problematic function call. 


\subsection{Comparison with State-of-the-art}
Finally, we compare our approach with the current state of the art approach for performance bug fixing for C\#: DeepDev-PERF ~\cite{FSEPerf}. DeepDev-PERF uses a sequence to sequence transformer model, BART~\cite{lewis2020bart} and finetunes it for the task of generating performance bugs. The model can output entire patches, as opposed to just completions, allowing it to generate fixes spanning the entire class. In contrast, our approach is intended to fix single method bugs only. Therefore, we only compare our model with DeepDev-PERF's ability to fix single method fixes, using the dataset they provide.

Table \ref{automated_eval_table_2} shows that RAPGen outperforms DeepDev-PERF both in automated and human evaluation metrics. This result is quite encouraging given that RAPGen only leverages our novel prompting approach in zero-shot, while DeepDev-PERF relies on expensive fine-tuning steps over a large dataset. Furthermore, it is also important to note that in DeepDev-PERF, the authors allow their model 2000 attempts for each example, and use the top-500 (based on average token likelihoods) for their evaluation, whereas, we only sample 100 completions per example. Despite the much fewer sampling attempts, our approach is still able to out-perform their finetuned model. One reason for this gap may be that our model includes information regarding buggy line, within the prompt, whereas the input to DeepDev-PERF doesn't localize the bug beyond just the buggy method. Therefore, their model may need to suggest many possible fixes before it finds the right one.

\section{In-The-Wild Evaluation}

To see the effectiveness of our approach in practice, we conducted an evaluation on existing .NET projects to verify the effectiveness of our approach in helping developers fix real world performance problems. 

\subsection{Experimental Setup}
\subsubsection{Finding real world problems to fix} We start with 50 codebases, internal to a large company, that are associated with existing services running in the cloud. Each of these services is being profiled by a profiler. The collected profiles are then analyzed by a commercial performance bug detection service that continuously analyzes profiler traces from a given application in the cloud and finds performance bottlenecks. 

Each bottleneck is of the form $(b, p)$, where $b$ in the bottleneck method that's usually an API method and $p$ the calling method from the user's codebase. We look at the insights generated by the performance bug detection service over CPU profiler traces. For each application, we find bottlenecks $(b, p)$ that meet the following criteria:
\begin{itemize}
\item The bottleneck takes up more than >5\% of the application's CPU. We set this threshold because we want to fix issues that take up a substantial amount of resources and are worth fixing. These are the issues that a developer would typically look into when improving the performance of their application.
\item $b$ is present in the RAPGen knowledgebase (Section \ref{kb})
\item A call to the method $b$ is present in the code for the parent method $p$. For example, if the bottleneck method $b$ is \texttt{Dictionary.ContainsKey}, we look for a call to \texttt{ContainsKey} in the parent method body. We need to confirm this because, this is often not the case when there is inlining of methods or the code uses syntactic sugar (e.g. use of special characters like "+" to perform \texttt{String.Concat} when concatenating strings).
\ref{index})
\end{itemize}

\subsubsection{Reaching out to developers} We take 10 random bottlenecks that meet the above criteria and each come from a different service. We then locate the code associated with each bottleneck in the corresponding codebase and use RAPGen to generate a fix. We communicate the generated fix to the engineers working on the project via email (Figure \ref{email}). In the email, we ask them to respond to the following questions regarding the fix:

\begin{itemize}
\item (Q1) Does the code change provide any performance improvement? (Yes / No)
\item (Q2) Did you need to make any significant modifications to the proposed fix? (Yes / No)
\item (Q3) How satisfied are you with the coding standards and practices used in the bug fix? (Very Satisfied / Somewhat Satisfied / Neither Satisfied or Dissatisfied / Somewhat Dissatisfied / Very Dissatisfied)
\item (Q4) How would you rate the overall quality of the code in the bug fix? (Excellent / Good / Average / Poor)
\label{survey_questions}
\end{itemize}

\subsection{Results}

We were able to secure responses to 8 of the emails i.e. a 80\% response rate. 7 of the 8 respondents confirmed that the changes would improve the performance of the code. 1 of the respondents rejected the fix. 6 in 7 respondents who accepted the fix confirmed that the changes could be accepted without any modifications. One of the respondents who accepted needed to make a minor change to the code. 

For the next question, we asked the developer to rate their satisfaction by picking from the following options: Very Satisfied, Somewhat Satisfied, Neither Satisfied or Dissatisfied, Somewhat Dissatisfied and Very Dissatisfied. 6 in 7 developers responded with \textit{Very Satisfied}. The final developer responded with \textit{Satisfied} because they had to make slight modifications to the code. As our final question, we asked the developer to pick from the following options: Excellent, Good, Average, Poor. 6 in 7 developers rated the overall quality of the fix as \textit{Excellent}, except for one developer who responded with \textit{Good}. Overall, this in-the-wild evaluation confirms the usefulness of our approach for fixing perf bugs in real life applications.

\section{Threats To Validity}
In this work, we focus only on the subset of performance issues caused by misuse of common .NET APIs. However, it's possible to have performance problems that are due to non-API related reasons~\cite{FSEPerf}, such as unnecessary computation in a loop that could be hoisted out or excessive allocations of a given type, etc. 
We also limit our scope to performance issues that can be fixed by changes to a single file and class. However, it's possible for fixes to span multiple classes, files or folders. This would require a much larger context window as we would need to pass code from multiple classes and the model would then decide which need to be modified in order to fix the issue. A concrete example of this could be a fix that requires making changes to a class that's used in multiple locations in a given code-base. The fix may require changes to not only the problematic class, but the usages as well. We plan to explore this in future work.

Additionally, we only explore fixing performance issues and assume that there already exists a way to identify expensive parts of a code-base, down to the exact line number. In .NET, this is indeed possible. One could use a profiler~\cite{app-insight-2019} and .NET's Source Link~\cite{sourcelink} technology, which adds metadata containing line number, commit, branch, etc. to a given profiler trace. We have not yet explored this for other languages like Java, python, etc., but we assume that similar technologies exist there as well. We leave this exploration to future work.

Finally, we were unable to run the code fixes suggested to the benchmark dataset. This is because many of the GitHub projects these examples were drawn from were not buildable, for variety of reasons such as, requiring proprietary packages to build, custom build steps or simply compiler errors. However, since our human evaluation showed that the fixes were equivalent to those proposed by the developer, we believe that our changes would result in performance gain as well. In practice, one could add an additional step to run benchmark tests to verify the change improves performance. Furthermore, the benchmark test could itself be generated using an LLM. We leave this exploration to future work. There already exists work that has already explored generating unit tests using these models ~\cite{Tufano2020UnitTC}.

\section{Related Work}
Next we discuss the work closely related to ours.

\subsection{Automated Program Repair}
Traditional automated program repair (APR) techniques usually require test suits or logic assertions as a specification of correctness. For example, GenProg \cite{genprog} uses genetic programming to generate code fixes that pass a supplementary suite of test cases. Pattern-based Automatic program Repair (PAR) \cite{auto_patch_gen} leverages manually created fix templates based on patterns in existing developer patches to guide the genetic search. 
Similarly, CapGen \cite{capgen} attempts to generate patches at finer granularity (AST node level) using context-aware prioritization of genetic mutation operations, and VarFix \cite{varfix} further extends GenProg by using variational execution \cite{variexec}. One of the major challenges with these APR approaches is overfitting \cite{ye2021comprehensive} and high maintenance cost. They also usually require a significant amount of effort on modeling program semantics. To fill this gap, our approach leverages prompt engineering on LLMs and therefore fixes a wide-range of performance bugs with low training and maintenance cost.

\subsection{Neural Code Fix Generation}
Recent advancements in deep learning, which automatically learn patterns from unstructured data, have driven a shift in APR techniques away from genetic programming and towards learning-based methods. Machine learning-based APR approaches leverage generative deep learning models that directly output patched versions of faulty code. For example, \cite{tufano2018} uses an RNN encoder-decoder model to generate fixes for bugs on GitHub projects. DeepDebug \cite{Drain2021DeepDebugFP} trained a backtranslation model to generate synthetic buggy versions of code to fine-tune a bug patch model. Stack traces for these generated bugs, as well as supplementary context through code skeletons, are used to finetune a bug patch generation model. 
Similarly, VRepair \cite{vrepair} trains a vanilla Transformer to automatically repair security vulnerabilities using transfer learning. The model is initially pretrained on bug-fix pairs from GitHub and fine-tuned on security vulnerability-fix pairs. 
VulRepair~\cite{vulrepair} builds on VRepair by replacing the vanilla Transformer with a T5 \cite{t5} architecture with BPE tokenization. 
TFix \cite{berabi2021tfix} is a transformer-based model that fixes bugs found with a static analyzer. More recently, studies have shown that Codex~\cite{Codex} can also generate patches to vulnerable code snippets \cite{codex_fix}. Similarly, \cite{codex_quixbugs} explored the viability of using Codex in a zero-shot setting to fix bugs in the QuixBugs benchmark \cite{quixbugs}. In this setting, Codex was indeed able to generate bug fixes, although fix performance was heavily dependent on the prompts fed to the model, which were created manually. While the majority of the work is focused on general bug fixing or vulnerability fixes, our work uses prompt engineering for fixing performance bugs specifically.


\subsection{Prompting and Prompt Engineering}
With the rise of LLMs prompt engineering ~\cite{liu2023pre} has emerged as a process to come up with the best prompt for a downstream task. A high quality prompt is often necessary to elicit the correct response from a language model. 
However, coming up with the optimal prompt is no trivial task~\cite{prompt_mining}. Prior work have explored various prompt engineering approaches including manually engineering suitable prompts \cite{petroni2019language, brown2020language}, few-shot prompting, and optimizing the ordering of prompt examples~\cite{lu2021fantastically} among others. Similarly, continuous prompts have emerged as an automated prompt learning approach ~\cite{li2021prefix, liu2021gpt, lester2021power}. Continuous prompts directly prompt the embedding space of the model and usually require labeled data. 
Our work uniquely contributes to prior work by introducing a new prompt generation method called RAPGen, which leverages a knowledge-base of instructions to guide the generation of the prompt.

\section{Conclusion}
In this work, we presented Retrieval Augmented Prompt Generation (RAPGen), a novel prompt engineering approach for fixing performance bugs using LLMs. RAPGen uses past performance fixes to construct a knowledge-base instructions encoding the changes made within these fixes. When given a code snippet with a performance issue and the corresponding expensive line, RAPGen leverages this knowledge-base to construct a prompt and generates fixes using LLMs in zero-shot. Being zero-shot and entirely prompt-driven enables RAPGen to leverage the capabilities of LLMs for Code without the cost of fine-tuning. Our extensive empirical evaluation shows that about 40\% of the fixes produced by RAPGen match the developer's fix and in $\sim$60\% of the cases, it is able to generate a fix that is equivalent or better than the developer fix. Our "in-the-wild" evaluation shows that the model is capable of producing fixes for performance problems in real world applications that are considered useful by developers. While we only explored our approach for the task of performance bug fixing, our knowledge-base creation and retrieval methods are not specific to performance bugs and can be extended to other types of bugs as well. We believe that our work uniquely contributes to the area of prompt engineering, specially for code-related downstream tasks.  

\bibliographystyle{IEEEtran}
\bibliography{references}

\end{document}